\documentclass[fleqn,usenatbib]{mnras}

\usepackage[T1]{fontenc}
\usepackage{ae,aecompl}

\usepackage{graphicx}	
\usepackage{amsmath}	
\usepackage{amssymb}	

\makeatletter
\newlength{\abovecaptionskip}%
\makeatother
\usepackage{threeparttable}%

\usepackage{newtxtext,newtxmath}

\title[VLA and ATCA monitoring of PKS~1830-211]{An improved time delay from VLA and ATCA monitoring of the gravitational lens system PKS~1830-211}

\author[A. D. Biggs]{
  A.~D.~Biggs$^{1}$\thanks{E--mail: andrew.biggs@stfc.ac.uk}
  \\
  $^{1}$UK Astronomy Technology Centre, Royal Observatory, Blackford Hill, Edinburgh EH9 3HJ \\
}

\date{Accepted XXX. Received YYY; in original form ZZZ}

\pubyear{2023}

\begin{document}
\label{firstpage}
\pagerange{\pageref{firstpage}--\pageref{lastpage}}
\maketitle

\begin{abstract}
We have measured a time delay of $25.3 \pm 2.0$~d (1-$\sigma$ confidence) in the Einstein ring gravitational lens system PKS~1830-211 from an analysis of archival VLA and ATCA monitoring data observed between 1997 and 2004. A small portion of the ATCA data was previously used to determine a time delay and our result is consistent with the previous value, but with an uncertainty that is smaller by more than a factor of two. The long time-baseline of the monitoring reveals that the flux density ratio is smoothly varying on a time-scale of years, an effect which we attribute to millilensing by massive objects ($\gg 1$~M$_{\sun}$) in the lensing galaxy. Image~A is unpolarized in the VLA monitoring, but VLBI observations show that this is partly due to beam dilution by an unpolarized counter-jet that is only present in that image. Based on the identification of this feature as a counter-jet, we conclude that its unexpected prominence in image~A is a consequence of lensing and that more detailed modelling is required in order to reconcile the VLBI morphology of each image.
\end{abstract}




\section{Introduction}
\label{sec:intro}

PKS~1830-211 is a gravitational lens system consisting of two images of a $z = 2.507$ quasar \citep{lidman99} and a prominent Einstein ring \citep{prameshrao88,subrahmanyan90,jauncey91}. Both images are extremely strong at radio frequencies, each having a flux density of several Janskys. The presence of both molecular \citep{wiklind96} and atomic absorption \citep*{chengalur99} strongly suggests that the lensing galaxy is a spiral and high-resolution optical data from the \textit{Hubble Space Telescope} have directly confirmed this \citep{lehar00,winn02a,courbin02}.

The molecular absorption system in the lensing galaxy at a redshift of $z=0.89$ \citep{wiklind96} is extremely rich, with many species having been detected, predominantly against image~B. In contrast, the atomic absorption (\ion{H}{i} and OH) is stronger in front of image~A, with an additional \ion{H}{i} absorption system present at a redshift of $z = 0.19$ \citep{lovell96,verheijen01}. See \citet{combes21} for an overview of the absorption systems in this lens system.

As well as giving the redshift of the lensing galaxy, the molecular absorption also allows a unique way of determining the time delay between the two lensed images. Using a single dish alone and thus without separating the two images spatially, it is possible to measure the continuum level and the depth of the absorption (the $\mathrm{J}=2-1$ transition of HCO$^+$) which, as noted, only occurs in front of image~B. From these two measurements, it is possible to determine the flux densities of the two lensed images and thence a time delay \citep[$24^{+5}_{-4}$~d --][]{wiklind01}.

Time delays have also been measured from unresolved monitoring of gamma-ray flares where, although the flaring activity from each image cannot be separated, auto-correlation techniques can extract the time delay signal \citep[e.g.][]{barnacka11}. This was clearly demonstrated for JVAS~B0218+357 \citep{cheung14} where the gamma-ray time delay is in agreement with the radio value \citep{biggs18}. However, in the case of PKS~1830-211 the situation is less clear, with studies arguing both for \citep{barnacka15} and against \citep{abdo15,abhir21} there being a time-delay signal in the data.

The only robust time delay resulting from data that resolves the lensed images was measured using the Australia Telescope Compact Array (ATCA) at 8.64~GHz \citep[$26^{+5}_{-4}$~d -- ][]{lovell98}. Earlier, an incorrect time delay \citep{vanommen95} was published using a limited amount of data from the Very Large Array (VLA), but three additional VLA monitoring campaigns have been undertaken and none published. \citet{muller23} recently presented a new time delay using the Atacama Large Millimetre/submillimetre Array (ALMA), albeit one utilizing a parameteric model of the source structure. Given the enduring interest in this source, we have analysed a robust subset of the VLA and ATCA data and present the results here.

The paper is organized as follows. In Section~\ref{sec:obs} we introduce the VLA and ATCA datasets and detail their calibration and the subsequent production of radio variability curves. We determine a time delay in Section~\ref{sec:delay} and discuss the results in Section~\ref{sec:discussion} where we also perform our own analysis of the \citet{wiklind01} data and present the first reliable VLBI imaging of this source at 8.5~GHz, including the distribution of polarized emission in the lensed source. Finally, we present our conclusions in Section~\ref{sec:conclusions}.

Regarding nomenclature, throughout this paper we refer to the lensed images using the traditional naming scheme i.e. A, B and C, where the former refers to the strongest. Many authors instead use a directional notation e.g.\ NW for A and SW for B.

\section{Observations and data reduction}
\label{sec:obs}

\begin{table*}
  \centering
  \caption{Summary of the three VLA monitoring seasons of PKS~1830$-$211 presented in this paper.}
  \label{tab:obs}
  \begin{tabular}{cccccc} \\ \hline
    Season & Project code & Dates & Configurations & Frequency bands (GHz) & Number of epochs \\ \hline
    1 & AH593 & 1996 Oct 12 -- 1997 May 16 & A, B & 8.4, 15 & 30 \\
    2 & AW576 & 2002 Jan 24 -- 2002 Sep 18 & A, B & 8.5 & 64 \\
    3 & AW607 & 2003 May 21 -- 2004 Jan 29 & A, B & 8.5 & 76 \\ \hline
  \end{tabular}
\end{table*}

\subsection{VLA}

The three seasons of VLA data are summarised in Table~\ref{tab:obs}. The first season, with VLA code AH593, was observed over a period of about eight months at two frequencies (8.4 and 15~GHz) in 1996/1997. The second and third seasons, AW576 and AW607, took place during 2002--2004 and aimed to measure the delays of three new lens systems found in a survey of the southern hemisphere \citep*{winn01b}, but also observed 1830$-$211 at 8.5~GHz. In all cases, observations were conducted in the two largest VLA configurations, A and B, resulting in an angular resolution at the lower frequency of $\sim$0.2 and 0.6~arcsec respectively. The sampling rate of the last two seasons was higher by a factor $>$2.

All observations were made in standard VLA continuum mode using two adjacent dual-polarization sub-bands, each with a bandwidth of 50~MHz -- full polarization products were produced by the correlator. Each epoch included at least one flux calibrator, a nearby phase calibrator and the lens itself. The first season used a single 3C source (3C~286 or 3C~48) as the flux density reference, whilst the Winn seasons used multiple Compact Symmetric Objects (CSOs), small ($\la$100~mas) lobe-dominated sources that have stable flux densities \citep{fassnacht01}.

Polarization calibration is possible for the AH593 data owing to the inclusion of a scan on the unpolarized calibrator OQ~208 which makes it possible to solve for the polarization leakage parameters (`D-terms'). Calibration of the polarization position angle (EVPA) can be accomplished with the same 3C source used for the flux-scale calibration. For the Winn data, the D-terms can be calculated using CSO scans as members of this class of source are generally unpolarized \citep{peck00}, but EVPA calibration is not possible as only a few epochs included a suitable calibrator.

The data were calibrated using \textsc{aips} \citep{greisen03} and a similar strategy was used for all epochs. Gain solutions were first found for all calibrators, using models for sources that were resolved, and the flux density of each found by comparing its solutions to those of the flux calibrator using \textsc{getjy}. For the Winn data, the best results were found using both PKS~J2130+0502 \citep{stanghellini97,dallacasa98} and J1400+6210 \citep{dallacasa13} at each epoch. Gains were then interpolated onto the lens from the phase calibrator scans that bracket it. Finally, polarization leakage was removed using the task \textsc{pcal} and the EVPA calibrated for the AH593 data only.

\begin{figure*}
  \begin{center}
    \includegraphics[width=0.45\linewidth]{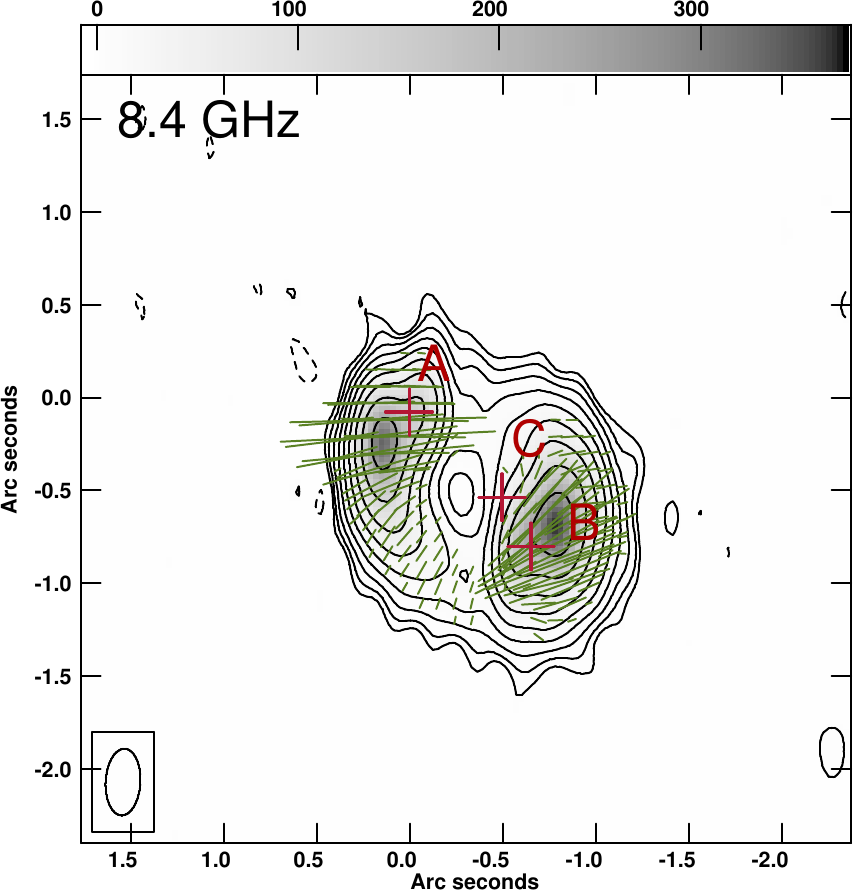}
    \includegraphics[width=0.45\linewidth]{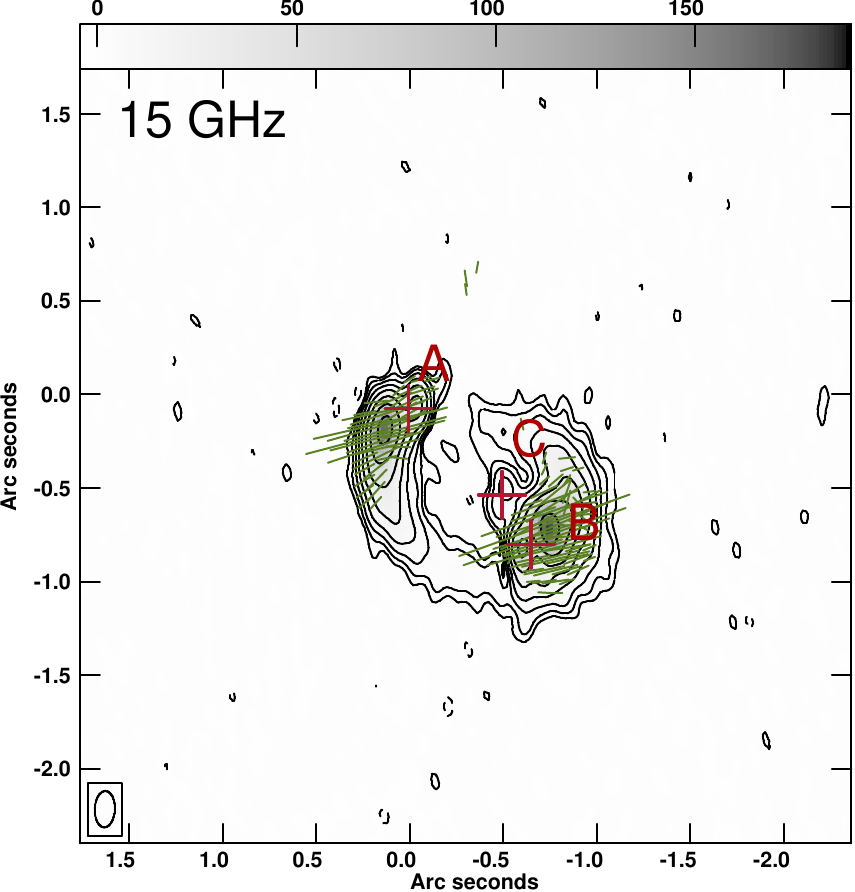}
    \caption{Total intensity maps (contours and grey scale) of PKS~1830$-$211 made using the AH593 data in which images A and B have been subtracted. Left: 8.4~GHz, Right: 15-GHz. The positions of the subtracted radio cores are marked with crosses, as is the position of the third image (C) as measured by \citet{muller20} which at 15~GHz can be seen to coincide with a feature labelled `E' by \citet{subrahmanyan90}. Also plotted are sticks indicating the orientation and relative magnitude of linear polarization. The restoring beams (bottom-left of each image) have dimensions of $360 \times 187$ and $194 \times 107$~mas$^2$ at 8.4 and 15~GHz respectively and contours are plotted at $-1$, 1, 2, 4. 8, etc. multiples of three times the rms noise ($\sigma_{\mathrm{8.4}} = 323$ and $\sigma_{\mathrm{15}} = 242~\mu$Jy\,beam$^{-1}$). The grey scale gives the image intensity in units of mJy\,beam$^{-1}$. Positions are offset from $18^{\mathrm{h}} 33^{\mathrm{m}} 39\fs9310, -21\degr 03\arcmin 39\farcs750$ (J2000) and have an astrometric accuracy of $\sim$10~per~cent of the synthesized beam-size -- this could account for the offset between image~C and the peak of `E'.}
    \label{fig:ring+pol}
  \end{center}
\end{figure*}

The flux density of the two lensed radio cores were calculated by fitting to the calibrated \textit{u,v} data using the Caltech \textsc{difmap} software \citep{shepherd97}. In doing this, it is necessary to account for the bright Einstein ring structure and we do this by following the same strategy used by \citet{biggs18} for JVAS~B0218+357. We first measure the flux densities of the lensed cores (modelled using a delta component) for the A-configuration epochs and with the shortest baselines ($<$400~k$\lambda$, where the ring dominates) excluded from the fit. These are then subtracted from the \textit{u,v} data and the epochs then combined into a single dataset and a map of the ring created. This process is then repeated in an iterative fashion with the improving ring map included in the \textsc{difmap} fit and with the \textit{u,v}-restriction eventually removed. As the EVPA is calibrated for the AH593 data, this process can be done for the Stokes \textit{I}, \textit{Q} and \textit{U} data and in Fig.~\ref{fig:ring+pol} we show final images of the ring at 8.4 and 15~GHz including EVPA vectors.

A final modelfit uses the best map of the ring to produce radio light curves of the two lensed cores in total flux density and, for AH593 only, the magnitude and position angle of polarization. When fitting the Stokes $I$ data, the model is optimized iteratively, interleaving model optimizations with self-calibration of the phases. As the source is very strong, we also perform a final self-calibration step that calculates a single amplitude correction for each antenna. This has little effect on the fitted flux densities, but does reduce the chi-squared as well as the noise in the residual map.

The error on the total flux densities is calculated from a quadrature sum of the noise in a naturally weighted residual map and a term representing the error in the setting of the flux scale. Based on the scatter in the flux densities of the CSOs \textit{not} used to set the flux scale, we estimate this to be 0.35~per~cent of the source flux density for AW576 and 0.45 for AW607. For AH593, we use a value of 0.8~per~cent. As the total flux density of each image is very high, the fractional error is the dominant term. The polarization errors are calculated using the formulation of \citet{wardle74} and include a term for a residual polarization leakage of 0.1~per~cent of the total flux density.

The final step is to remove unreliable epochs. Those are identified on the basis of a relatively poor fit to the \textit{u,v} data (high reduced chi-squared) or a relatively high rms in the residual map. In addition, the large number of stable flux calibrators present in the Winn data allows us to identify those with poor flux calibration. One AW576 B-configuration epoch was flagged due to the hour angle of the observation resulting in such poor \textit{u,v} coverage that the modelfit was very unreliable. Three, eleven and thirteen epochs were flagged in each season respectively.

\subsection{ATCA}

The ATCA monitoring data presented by \citet{lovell98} were actually part of a much larger campaign (comprising 123 epochs) conducted between January 1997 and August 1999 (project code: C618). To these we add 12 epochs observed as part of a monitoring campaign unrelated to lensing \citep[project code: C540;][]{tingay03} giving a total of 135 epochs. Although both campaigns observed in multiple frequency bands, only the highest (8.64~GHz) gives sufficient angular resolution to resolve the two lensed images. The total bandwidth was 128~MHz.

In both cases, each epoch consisted of short observations of 1934-638 to set the flux scale and J1924-2914 for phase-calibration purposes. The total time spent observing 1830-211 varied greatly. For the Tingay epochs, each epoch only consisted of a few minutes on source whilst for the Lovell data, the on-source time varied between 10~min to several hours. The Lovell data provide full polarization information, but as most epochs have insufficient parallactic angle coverage, we do not derive any polarization results.

The ATCA data were calibrated using \textsc{CASA} \citep{casa22} following standard recipes. The data were first converted from the native RPFITS format into measurement sets using \textsc{importatca}. Bandpass calibration for all sources was performed using 1934-638 and this source was also used to set the flux scale, using the Perley-Butler 2010 standard. Gain solutions (amplitude and phase) were found for all sources and those of J1924-2914 interpolated onto the 1830-211 data. Light curves were derived for the ATCA data in exactly the same way as for the VLA using \textsc{difmap}.

\begin{figure*}
  \begin{center}
    \includegraphics[width=0.9\linewidth]{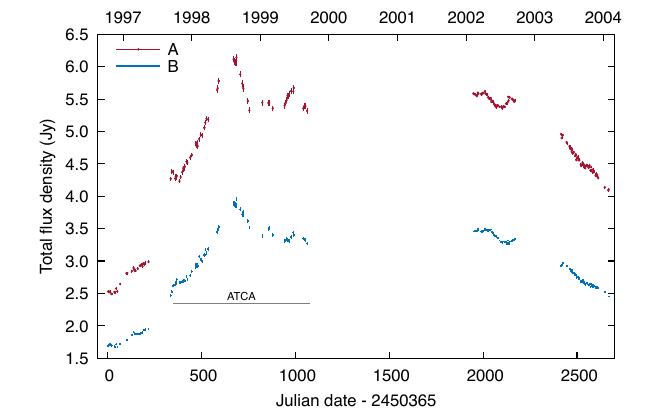}
    \caption{VLA and ATCA radio light curves of PKS~1830$-$211 at 8.4-8.6~GHz. The error bars are in many cases smaller than the plotting symbol.}
    \label{fig:vc_x}
  \end{center}
\end{figure*}

\begin{figure*}
  \begin{center}
    \includegraphics[width=0.45\linewidth]{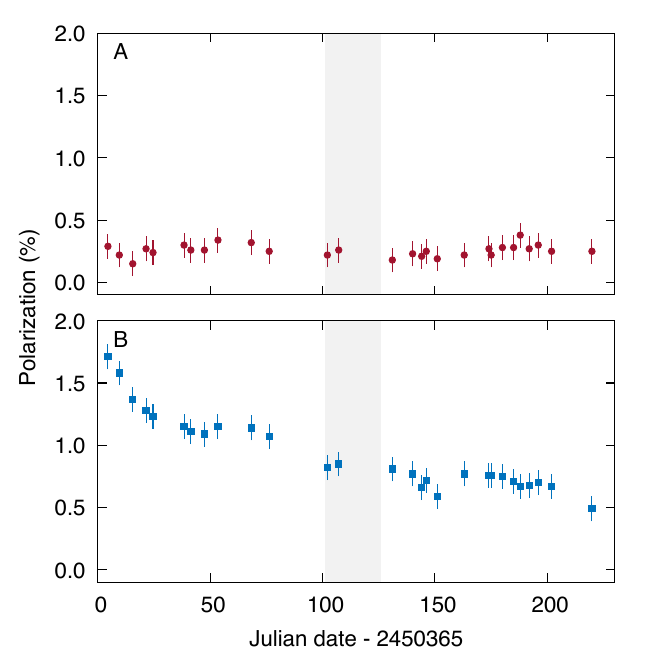}
    \includegraphics[width=0.45\linewidth]{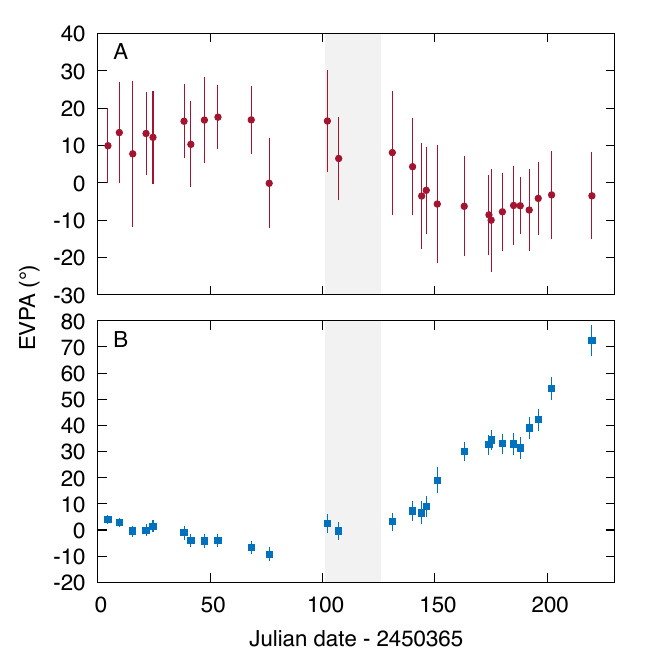}
    \caption{VLA 8.4-GHz percentage polarization (left) and EVPA (right) plots of PKS~1830$-$211 from Season~1. The polarization of image~B declines by about a factor of two over the course of the monitoring while the EVPA shows a rotation of about 90\degr. The data for image~A, in contrast, have a very low SNR and show no sign of variability.}
    \label{fig:vc_x_pol}
  \end{center}
\end{figure*}

As the ATCA monitoring data do not contain observations of a second non-variable source for the purposes of measuring the flux-calibration accuracy, we start with the value of 0.7~per~cent measured by \citet{winn04a} for a very similar monitoring campaign which included observations of a CSO. However, our calibration will be less accurate as we are not able to calibrate out the effects of the antenna gain changing with elevation. Assuming that the gain-elevation curves used by \citet{winn04a} (kindly provided by H. Bignall) are representative of the magnitude of correction that would be required for the Lovell data, we calculate an additional rms scatter of 0.4~per~cent. When combined in quadrature with the gain-elevation-corrected value of 0.7~per~cent, this gives a total flux accuracy for the Lovell data of 0.8~per~cent.

Epochs where poor weather had rendered the data unreliable were identified on the basis of their data weights which, for those with obvious flux-calibration errors, were abnormally low. We flagged 13 epochs on this basis, an additional two due to antenna CA06 being absent (without which the lens images are not resolved) and another due to the lens having being observed at too high an hour angle. We also disregard a short period of 11 epochs observed four months before the rest.

Ultimately, however, another effect has led us to remove many more epochs. ATCA is an East-West array and consists of only six antennas with a maximum baseline length of $\sim$6~km, approximately half that provided by the VLA's B-configuration and just about sufficient to resolve the two lensed images of 1830-211. Many of the ATCA configurations do not contain baselines long enough to constrain the A/B flux ratio (the input and output values are the same) and therefore we only consider data taken using a `6A', `6B', etc. or `375' configuration. All in all, this leaves a total of 63 epochs.

\section{Time-delay analysis}
\label{sec:delay}

The 8.5-GHz total flux density data is shown in Fig.~\ref{fig:vc_x} where it can be seen that the source varies significantly throughout the monitoring period. The variations are rather smooth in appearance, there being little sign of variability on time-scales of a month or less. The polarization data are shown in Fig.~\ref{fig:vc_x_pol} for Season~1 only. Large variations are seen in image~B which declines in magnitude of polarization by a factor of two and undergoes a 90-degree rotation of the EVPA. Image~A, however, appears to be undetected. The relatively constant value of the EVPA here suggests that what is actually being measured is residual emission from the Einstein ring.

To determine the time delay, we restrict ourselves to the ATCA data and the final two VLA seasons. The first VLA season is of poorer quality than the other two and has a large gap about halfway through. The ATCA data, although also containing gaps, display the largest variations and cover a long time period of about two years. We shall determine the time delay for the ATCA and Winn data separately and together, using two techniques: chi-squared minimisation (CSM) and the dispersion statistic of \citet{pelt96}.

The CSM method first shifts one curve by a trial time delay ($\tau$) and flux density ratio ($r$) and calculates the reduced chi-squared by comparing each epoch in one curve with an interpolated value in the other. In Equation~\ref{eq:csm}, $\nu$ is the number of degrees of freedom and interpolated values are shown with a tilde.
\begin{equation}
  \label{eq:csm}
  \chi_{\nu}^2(\tau) = \frac{1}{\nu} \sum_i^N \frac{(f_{\mathrm{A}} - r \tilde{f_\mathrm{B}})^2}{\sigma^2_{\mathrm{A}} + r^2 \tilde{\sigma}^2_{\mathrm{B}}}
\end{equation}
No extrapolation is done and so as the time delay is increased, the overlap region shrinks and fewer pairs, $N$, are included in the statistic. This is calculated twice at each delay, the second time shifting the two curves by the negative of the time delay and interpolating in image A -- the two values are then averaged.

Interpolation does not feature in the Pelt method which instead forms a dispersion, $D$, by comparing neighbouring measurements in a combined time-ordered light curve, $C$, formed by shifting one image by $\tau$ and $r$. The $D^2_3$ variant (Eq.~\ref{eq:pelt}) forms a weighted squared difference from pairs of measurements as
\begin{equation}
  \label{eq:pelt}
  D^2_3 = \frac{\displaystyle\sum_{l=1}^{N}\sum_{m=l+1} S_{l,m} W_{l,m} G_{l,m} \left(C_{m} - C_l\right)^2}{2 \displaystyle\sum_{l=1}^{N}\sum_{m=l+1} S_{l,m} W_{l,m} G_{l,m}}.
\end{equation}
$W$ is a weighting term formed from a combination of the uncertainties on each measurement and $G$ takes a value of 1 if the measurements are from different images and 0 otherwise. $S$ works similarly to $G$ and determines where measurements are in relation to each other in time. If they occupy consecutive positions in the time series and are separated in time by less than $\delta$, $S=1$ and 0 otherwise. We have used $\delta = 10$~d to avoid measurements on opposite sides of gaps in the ATCA data from contributing to the dispersion, although values in the range 5--20~d result in the same delay.

We also use the $D^2_{4,2}$ variant which calculates the dispersion for all measurement pairs, the contribution of each to the dispersion being regulated by a modified version of $S$ which varies smoothly between 0 and 1 according to $\delta$ which now takes on the role of a so-called decorrelation length:
\begin{equation}
  \label{eq:delta}
  S_{l,m} = \frac{1}{1 - \frac{|t_l - t_m|}{\delta}}.
\end{equation}
We set $\delta$ to the same value as the cut-off in the $D^2_3$ variant i.e.\ 10~d.

\begin{figure*}
  \begin{center}
    \includegraphics[width=0.9\linewidth]{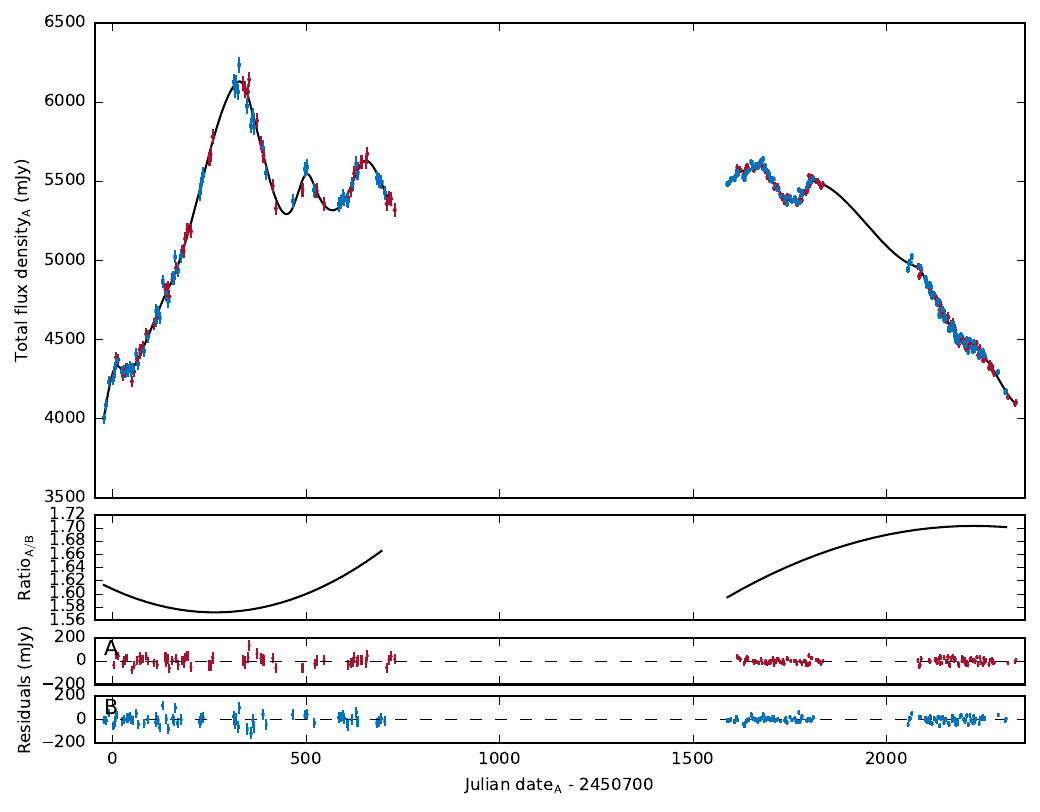}
    \caption{Results of the CSM time-delay analysis. Top panel: Combined total flux density radio light curve after removal of the time delay (25.3~d) and flux density ratio. The solid line is a cubic spline fit to the data. Middle panel: polynomials representing the time-varying flux density ratio between A and B. Bottom panel: residuals for each image around the spline.}
    \label{fig:vc_x_comb}
  \end{center}
\end{figure*}

It is necessary to make the flux density ratio a function of time as fitting a constant gives a poor agreement between the two images. We have therefore modelled this as a second-order polynomial, the lowest order that gives an acceptable by-eye fit and a reduced chi-squared of $\sim$1. Separate polynomials are used for the ATCA and Winn data.

The results of performing the time-delay analysis on the combined ATCA and Winn dataset using the CSM method are shown in Fig.~\ref{fig:vc_x_comb} where we show the combined light curve together with the fitted flux-density-ratio polynomials. Also shown is a cubic-spline fit to the combined light curve and the residuals around this.

\begin{table}
  \centering
  \caption{Time delay (of B relative to A) and associated 1-$\sigma$ uncertainty of PKS~1830-211 as measured using the CSM and Pelt techniques. For each Pelt dispersion variant we set $\delta = 10$~d. All delays are consistent at 1~$\sigma$.}
  \label{tab:delay}
  \begin{tabular}{cccc} \\ \hline
    Dataset & Method & $\chi^2_{\nu}$ & $\tau_{\mathrm{B-A}}$ (d) \\ \hline
    ATCA & CSM & 1.02 & $23.2 \pm 3.2$ \\
    & Pelt ($D^2_3$) & & $26.0 \pm 2.6$\\
    & Pelt ($D^2_{4,2}$) & & $23.0 \pm 3.0$ \\ \hline
    Winn & CSM & 0.94 & $25.4 \pm 2.4$ \\
    & Pelt ($D^2_3$) & & $25.9 \pm 2.7$ \\
    & Pelt ($D^2_{4,2}$)& & $24.0 \pm 1.7$ \\\hline
    Both & CSM & 0.96 & $25.3 \pm 2.0$ \\
    & Pelt ($D^2_3$) & & $25.9 \pm 2.6$ \\
    & Pelt ($D^2_{4,2}$)& & $24.0 \pm 1.6$ \\ \hline
  \end{tabular}
\end{table}

Uncertainties on the time delay are determined by creating multiple realizations of the A and B light curves and recalculating the time delay and flux density ratio for each. The basis for this procedure is the cubic spline that represents the intrinsic variability of the source. This is sampled using the observed epochs, but for each image we shift these by a randomly chosen offset. In this way, the time delay of each simulated light curve can differ by up to 3~d from that measured using the real data.

Once simulated versions of the intrinsic light curves have been created, each measurement must then be perturbed by some estimate of that epoch's uncertainty. As there is no sign of independent variability in each image other than the long-term variation in the flux-density ratio that we model with a polynomial, it should only be necessary to perturb each measurement by a random value taken from a normal distribution with a standard deviation equal to the epoch's uncertainty. This leads to a goodness-of-fit ($\chi^2_{\nu}$ or dispersion) that is somewhat better than the original and so, to be conservative, we increase the uncertainties by between 5 and 10~per~cent to make them match.

We create 2000 realizations of the simulated light curves and from these determine the median time delay and the upper and lower 1-$\sigma$ bounds bounds containing the inner 68.3~per~cent of the results. The systematic shift represented by the median is then added in quadrature to a $\pm$0.5~d error representing the fact that delays tend to bunch in clusters separated by 1~d, a consequence of epochs typically being scheduled at intervals that are an integer multiple of 24~h \citep{biggs21}. The total systematic error is then added in quadrature to the random error defined by the largest of the 1-$\sigma$ bounds to form the final error estimate.

The final delays and their uncertainties are shown in Table~\ref{tab:delay}. The ATCA-only data produce the largest spread of delays (23--26~d) but also the largest delay uncertainties, as expected given the larger flux density measurement errors and the gaps in the light curves. The delays measured using the Winn data are slightly more consistent (24.0--25.9~d) and change little when the ATCA data are also included. However, adding the ATCA data reduces the uncertainty in the delay and thus we assume that the combined dataset gives the most accurate measurement of the time delay. As there is little to choose between the CSM and Pelt results, we favour the CSM value as it involves no free parameters. We therefore take the delay in 1830-211 to be $25.3 \pm 2.0$~d.

\section{Discussion}
\label{sec:discussion}

\subsection{The time delay and prospects for measuring $H_0$}
\label{sec:h0}
  
\begin{figure}
  \begin{center}
    \includegraphics[width=0.9\linewidth]{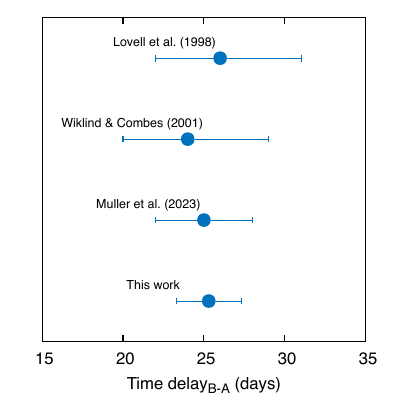}
    \caption{Comparison of the time delays measured by previous radio or mm monitoring campaigns.}
    \label{fig:all_delays}
  \end{center}
\end{figure}

Our time delay measurement from the combined ATCA+VLA monitoring data is shown in Fig.~\ref{fig:all_delays} together with the delays measured from other radio and millimetre monitoring campaigns. It can be seen that the new measurements are consistent with the previous estimates, although the uncertainty is smaller. The time delay in the radio and millimetre regimes seems therefore to be constrained to an accuracy of better than 10~per~cent.

For the purposes of lens modelling, the position of the third image (C) was recently confirmed using ALMA imaging \citep{muller20} and the Einstein ring is a plentiful source of constraints that has not yet been properly exploited. The LensClean algorithm was first demonstrated on PKS~1830-211 \cite{kochanek92}, but further developments of this method \citep*{ellithorpe96,wucknitz04a} have so far only been applied to other lens systems containing rings. More recently, \textsc{pyautolens}\footnote{https://github.com/Jammy2211/PyAutoLens} \citep*{nightingale21,nightingale18,nightingale15} has been made available to the community which uses a development of the semi-linear inversion method of \citet{warren03} to optimize a lens model and reconstruct the source and which, most importantly for 1830-211, can work with radio interferometric visibility datasets.

The main uncertainty in the lens model is the number of galaxies taking part in the lensing. As mentioned in Section~\ref{sec:intro}, a second absorption system is present at $z = 0.19$ and this seems most likely to be identified with a blue galaxy 2.5~arcsec to the south of the lens system. The most convincing case for this interpretation was given by \citet{winn02a} who noted a single counterargument, namely that one would expect the \ion{H}{i} absorption to be of equal strength in front of each image given that they are approximately equidistant from the galaxy. The discovery paper for this absorption system \citep{lovell96} indeed seemed to show that the absorption was much stronger in front of the NE image, but high SNR data taken with the VLA together with the Pie Town antenna unequivocally show that the absorption column in front of each image is in fact very similar \citep{verheijen01}. Both \citet{lehar00} and \citet{muller20} consider that if this galaxy does indeed lie at $z = 0.19$, it contributes very little to the lensing of the quasar.

This leaves the mass distribution within the Einstein ring that is responsible for the main lensing effect as the main uncertainty in the lens model. \citet{winn02a} favoured a scenario whereby the spiral is the only lensing galaxy, but \citet{courbin02} and \citet{meylan05} have detected a second extended object close to the centre of the spiral galaxy. Whether this is part of the spiral galaxy at $z = 0.89$ or a separate galaxy at an unknown redshift remains unclear.

\subsection{The changing flux density ratio}

One finding from the time-delay analysis is that the flux density ratio between A and B is changing with time. Temporal changes in the flux density ratio of this source have been noted before \citep[e.g.][]{prameshrao88,muller08,muller23}, but never from data with such high time sampling.

The evolution of the flux density ratio is extremely smooth. As we have seen, for the ATCA and Winn data separately it is possible to model this using a second-order polynomial. The reduced chi-squared after removal of the flux density ratio and time delay is very close to 1, indicating that there are no significant deviations from the smooth model. For both the ATCA and Winn data, the ratio changes by a maximum of $\sim$6~per~cent with an average value of $\sim$1.64. There is no indication of microlensing i.e.\ independent variations on time-scales of days or weeks such as has been confirmed as occurring in the lens system CLASS~B1600+434 \citep{koopmans00b,biggs21}.

As the changing flux density ratio is caused by independent variability in the two lensed images (it is impossible to know in which one) it indicates the presence of some form of propagation effect external to the lensed source. Two possibilities exist -- either some form of scintillation in the ionized ISM of our galaxy or an additional lensing effect caused by motion of a jet component in the source i.e.\ millilensing.

The former possibility seems unlikely. Due to its proximity to the Galactic centre ($l=12\degr, b=-6\degr$), VLA observations of 1830-211 should be subject to strong scattering as the transition into the weak regime is expected to lie at $\ga$40~GHz \citep{walker98,walker01}. Strong scattering comes in diffractive and refractive varieties and only the latter (refractive interstellar scattering or RISS) can produce variations with a time-scale of days or more \citep{narayan92}. However, RISS is not capable of producing correlated variability with a time-scale of years, even in the Galactic plane -- see \citet{rickett90} and references therein.

As gravitational lensing is achromatic, one might expect to see similar variations at different observing frequencies if this were the reason for the varying flux density ratio. For an extended source, the observed variability is the result of an average of the caustic pattern over the angular size of the source and, as radio source sizes are smaller at higher frequencies, the magnitude of variability should increase with observing frequency \citep{koopmans00b}. Although we have data at only a single frequency, \citet{wiklind01} refer to a linearly changing flux density ratio in their multi-season, single-dish monitoring campaign conducted at a frequency of 100~GHz.

\begin{figure*}
  \begin{center}
    \includegraphics[width=0.9\linewidth]{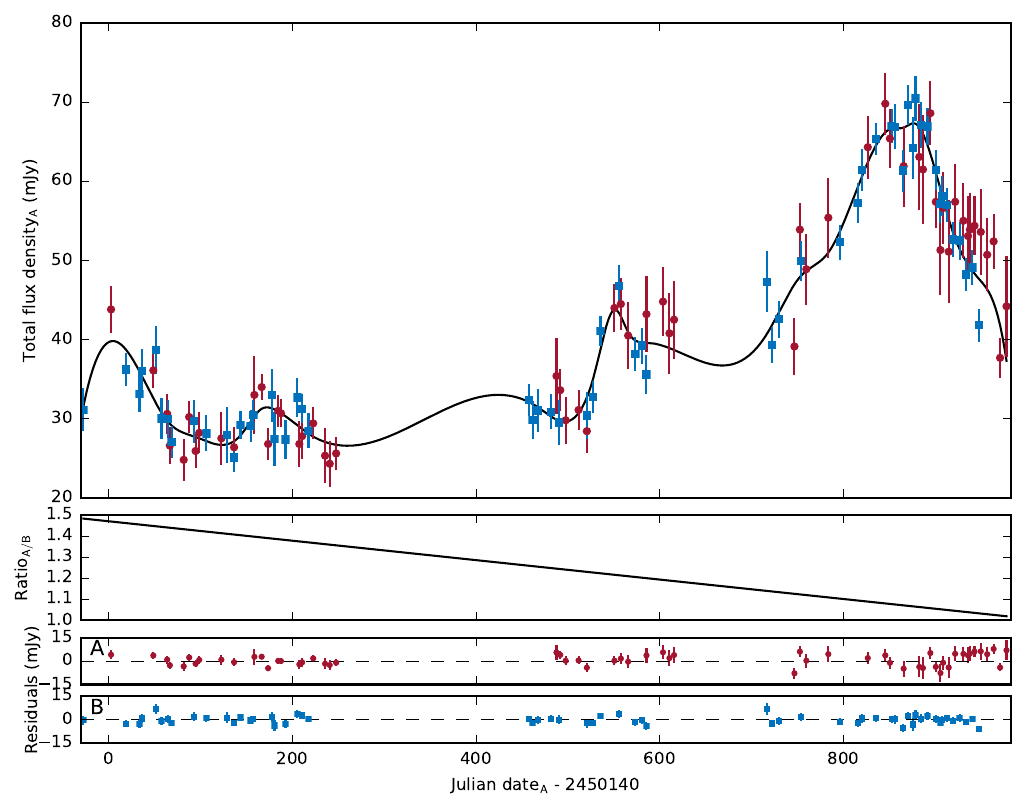}
    \caption{Results of the CSM time-delay analysis for the 100-GHz data of \citet{wiklind01}. Top panel: Combined total flux density radio light curve after removal of the time delay (24.4~d) and flux density ratio. The solid line is a spline fit to the data. Middle panel: linear function representing the time-varying flux density ratio between A and B. Bottom panel: residuals for each image around the spline. The prominent peak at the end of the monitoring corresponds to the maximum seen during the ATCA monitoring.}
    \label{fig:vc_3mm_comb}
  \end{center}
\end{figure*}

As no more details were given in the original paper, we have extracted the data from the PostScript of their fig.~1 as stored in the arXiv repository. This gives the flux density of the continuum and depth of the HCO$^+$~(2-1) absorption line from which it is possible to construct light curves for A and B, assuming that the absorption is only happening in front of image B. A CSM time-delay analysis with the flux density ratio modelled as a simple linear gradient is shown in Fig.~\ref{fig:vc_3mm_comb}. To the quoted precision, our delay is the same ($\Delta\tau_{\mathrm{AB}} = 24.4$~d cf. 24) and we see that the variation in the flux density ratio is much greater than seen in the VLA data, decreasing by a factor of 1.5 over the course of about three years.

Given that the external variability is much larger at 100~GHz, we ascribe its origin to millilensing of a superluminal jet component as it moves across the caustic structure associated with massive objects, these most likely being located within the lensing galaxy. \citet{koopmans00b} performed an in-depth analysis of radio microlensing in the lens system CLASS~B1600+434 and concluded that the variations seen in one image only were due to compact objects in the lensing galaxy with a mass of $\ga$0.5~M$_{\sun}$. For the much longer time-scale of the variations seen in 1830-211, the mass of the compact objects must be correspondingly larger as the Einstein radius of a lens increases with mass.

A good example of how massive substructures can cause the flux of a background jet component to change on a time-scale of years is given by the so-called symmetric achromatic variability (SAV) observed in the enigmatic radio source PKS~1413+135 \citep{vedantham17,peirson22}. Although the SAV in this source is modelled using binary lenses (in order to reproduce the distinctive double-horned variability pattern), the model light curves away from the peaks in the flux density illustrate well the sort of long-term variability that millilensing can produce.

\subsection{The non-detection of polarization in image A}

A surprising result from the monitoring was the non-detection of polarization in image~A over the full duration of the monitoring. That image~A seems to be less polarized than B has been noted since its discovery \citep{prameshrao88,subrahmanyan90}, but the apparent permanent nature of this at 8.4~GHz points to significant depolarization, perhaps due to differential Faraday rotation across the image which, when averaged over the VLA synthesized beam, reduces the polarization to below the detection threshold. To investigate this, we have analysed a Very Long Baseline Array (VLBA) 8.4-GHz observation from 1995 July 7 which has a synthesized beam area about $10^4$ times smaller than that of the VLA in A-configuration.

\begin{figure*}
  \begin{center}
    \includegraphics[width=0.45\linewidth]{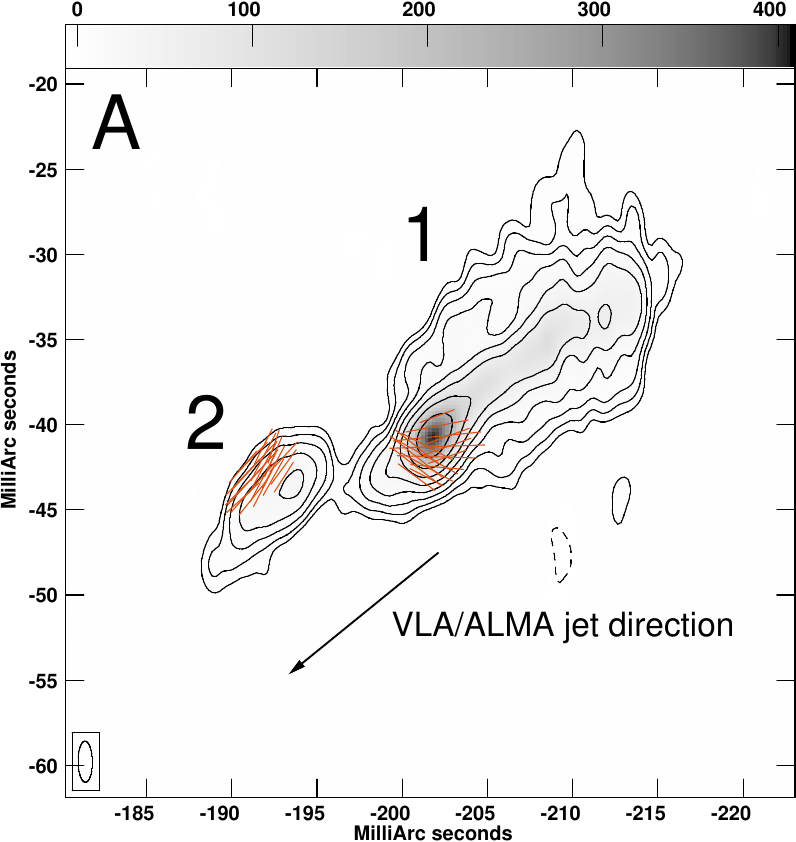}
    \includegraphics[width=0.45\linewidth]{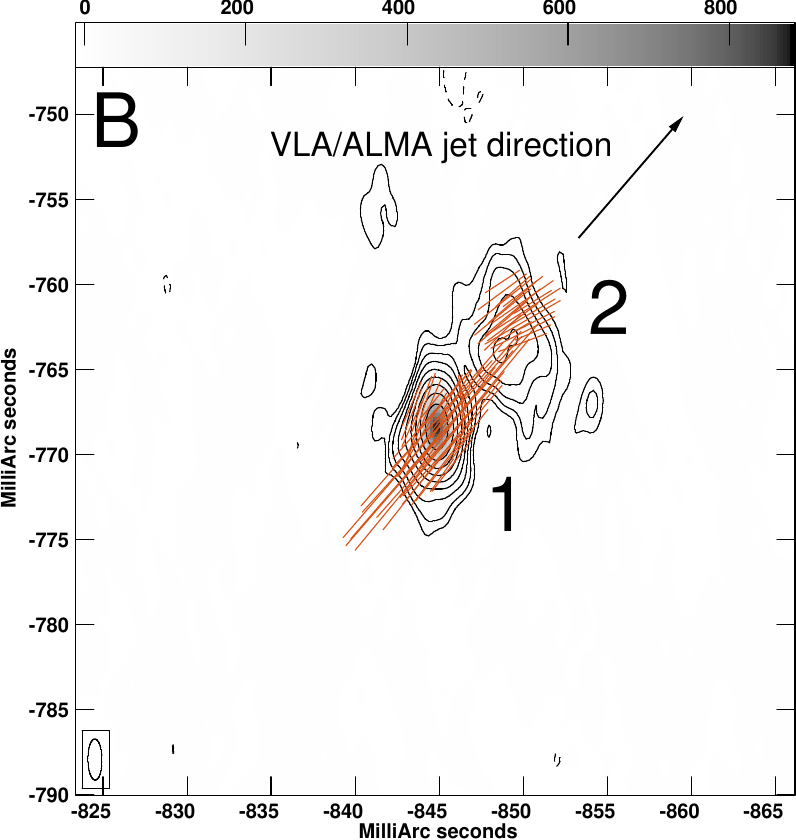}
    \caption{VLBA 8.4-GHz maps of image~A (left) and B (right) of PKS~1830$-$211 as observed on 1997 July 7. The synthesized beam has dimensions of $3.9 \times 1.3$~mas$^2$ and contours are plotted at $-1$, 1, 2, 4, etc. multiples of the rms noise (0.62~mJy\,beam$^{-1}$). The grey scale has units of mJy\,beam$^{-1}$ and sticks show the magnitude and orientation of polarization where a length of 5~mas corresponds to a polarized flux density of 3.33~mJy\,beam$^{-1}$. The source consists of two components -- a modestly polarized core (1) and a more highly polarized jet component (2) which points along the direction of the jet seen in lower-angular-resolution imaging. Component A1 seems to consist of an unpolarized, highly magnified counter-jet which may be entirely absent in B. The only evidence for a counterpart to this in B is a weak bulge to the south-east of the core.}
    \label{fig:vlbi}
  \end{center}
\end{figure*}

The VLBA data were taken as part of the observing programme BH010 (PI: J. Hewitt) and include 7 scans of 1830-211 spread over a period of 4~h for a total observing time of about 44~min. The bandwidth was $2 \times 8$~MHz per polarization and the correlator produced full polarization products. The polarization D-terms calibrator was 3C~84 and a single scan of J1310+3220 was used to set the absolute value of the EVPA. The data were calibrated and imaged in \textsc{aips} following standard procedures. Amplitude calibration was performed by applying measured system temperatures and a priori gain curves with each source fringe-fitted thereafter. For 1830-211, this was initially performed using a CLEAN component model made from VLA data, but this was iteratively improved upon using imaging and self-calibration of the VLBA data. Once satisfactory solutions were found, the data were further imaged and self-calibrated, firstly in phase and later in amplitude as well. No time or frequency averaging was performed, thus reducing any smearing effects to a minimum.

The final maps are shown in Fig.~\ref{fig:vlbi} and reveal that image~A \textit{is} polarized and that its polarization morphology is very similar to that of B i.e.\ there are two separate polarized components. In total flux density, however, the two images look very different with A containing a prominent extension to the north-west that is unpolarized.

B looks like a canonical core-jet with a small bulge approximately south-east (position angle of 120\degr\ -- North through East) of the compact core. The main jet component lies in the direction of the arcsec-scale jet seen most clearly in ALMA imaging \citep{muller20} and thus the weak emission on the other side of the core is likely to be associated with a counter-jet. The core and jet polarizations are $\sim$2 and 12~per~cent respectively. In Fig.~\ref{fig:vlbi} we have labelled these components B1 and B2 and also indicated the approximate direction of the jet as seen in VLA and ALMA images. We have estimated the flux density of the weak counter-jet by fitting a Gaussian to B1. The residual peak flux density is 26~mJy\,beam$^{-1}$ which gives a very approximate value for the strength of the counter-jet, just 1.5~per~cent of the flux density of the core.

For A, based on the total flux density imaging alone, it is clear that the most easterly component is the counterpart to the jet component B2 as they have a similar size, surface brightness and polarization (magnitude and EVPA) -- we therefore label this as A2. The polarization imaging also allows us to identify the peak in A1 as the counterpart to the radio core seen in B. The EVPA is very different, but this is to be expected for an image of the core of the radio source which is known to be variable (Fig.~\ref{fig:vc_x_pol}).

The unpolarized extension to the core of A is presumably the reason why no polarization is detected from this image in the VLA monitoring i.e.\ the polarization is diluted when averaged over the large VLA beam. Averaging the emission in the VLBI map of image~A would give a maximum polarization of only $\sim$0.3~per~cent if the EVPA of each component was aligned, and less otherwise. This is similar to the baseline level of polarization detected in image~A which indicates the 1-$\sigma$ noise level of our VLA monitoring.

Where is the unpolarized extension in B? Using the model of \citet{nair93}, \citet{garrett97} found that it should be clearly visible and attributed its absence to the additional lensing action of the foreground galaxy which hosts the $z=0.19$ \ion{H}{i} absorption system. This galaxy is now believed to have little influence on the lensed images \citep{lehar00,muller20}, but adaptive-optics imaging \citep{meylan05} has more recently revealed an additional extended component within the Einstein ring. The exact nature of this object and its redshift remain unknown and thus its contribution to the lensing is unknown. Furthermore, its location near the centre of the lensing mass argues against it only affecting image~B.

Alternatively, we should consider that this feature has not been extinguished in image~B, but somehow amplified in A. Whatever the reason for the prominent north-west extension in image A, it lies on the other side of the core to the known direction of the jet seen in ALMA and VLA imaging and as such must be considered a counter-jet. Counter-jets are, however, rarely seen in VLBI images of core-dominated radio sources as they should be receding from the observer and therefore subject to Doppler dimming. With this in mind, it is image~A that confounds our expectations as it has a counter-jet that is more prominent than the jet.

It might be tempting to ascribe the absence of the prominent counter-jet in B as being a consequence of the scatter-broadening that is known to affect this image \citep{jones96}. However, given that B2 is clearly detected with a surface brightness similar to A2, it seems unlikely that a feature located $\sim$10~mas away would be completely extinguished by the same effect. Therefore, we conclude that the inconsistent morphologies of A and B are a consequence of additional magnification of the source's counter-jet in image~A above what would be expected from a simple lens model.

\section{Conclusions}
\label{sec:conclusions}

We have performed an analysis of archival VLA and ATCA radio data with the main goal of improving the time delay between the two lensed images of PKS~1830-211. Whilst not considered to be a good candidate for $H_0$ determination, improvements in the lens model could stem from the now known position of the third image \citep{muller20} and the application of modern algorithms to the brightness distribution of the radio Einstein ring. Aside from lensing, the time delay remains relevant in terms of measuring possible offsets between the radio and gamma-ray emitting regions \citep[e.g.][]{barnacka15}. Our best estimate of the time delay is $25.3 \pm 2.0$~d, consistent with the value of \citet{lovell98}, but with the uncertainty reduced by a factor of two.

The flux density ratio was found to be slowly varying with time, by a maximum of $\sim$6~per~cent on a time-scale of a few years. A similar effect has been observed in the lens systems CLASS~B1600+434 \citep{biggs21} and CLASS~B1608+656 \citep{fassnacht02} and might be observed in more lens systems if they were monitored for long enough. A reanalysis of the 1830-211 single-dish monitoring data \citep{wiklind01} reveals considerably larger variations at $\sim$100~GHz which points to a lensing origin for the external variability. As the time-scale of the variations is much longer than that of the microlensing observed in CLASS~B1600+434 \citep{koopmans00b}, we conclude that the flux density of one or both of the images in 1830-211 is subject to millilensing by a fast-moving jet component moving behind relatively massive ($\gg$1~M$_{\sun}$) objects located in the lensing galaxy.

The lensed source has long been known to be polarized and yet no polarization was detected from image~A during the VLA monitoring. VLBI maps have confirmed that the polarization of image~A is very similar to that of B on milliarcsec scales, whilst the total flux density is surprisingly different, there being a large unpolarized extension to A. Whilst this feature naturally explains the lack of polarization in the VLA monitoring as being a consequence of beam dilution, why this is only present in image~A is not clear. \citet{garrett97} have shown that, based on a model of the lensing mass, it should be visible in both. However, whilst these authors consider that a perturbation to the lens model has rendered it undetectable in B, we conclude the opposite i.e.\ that a faint counter-jet in the lensed source has been magnified in image~A by much more than would be expected given a simple model for the lensing mass.

Finally, we note that our VLBI maps look very different to those of \citet{guirado99}, the most notable difference being the complete absence of their component NEX3 from our maps. The datasets were observed only ten days apart which rules out source variability as the cause of the discrepancy, although multi-epoch data \citep{garrett97} have in any case shown that A does not vary on a time-scale of about a year at 15~GHz. We have performed our own analysis of the \citeauthor{guirado99} data and find that A and B look very similar to the maps presented here. The new maps will appear in a future publication.

\section*{Acknowledgements}

The National Radio Astronomy Observatory is a facility of the National Science Foundation operated under cooperative agreement by Associated Universities, Inc. The Australia Telescope Compact Array is part of the Australia Telescope National Facility (grid.421683.a) which is funded by the Australian Government for operation as a National Facility managed by CSIRO. We acknowledge the Gomeroi people as the traditional owners of the Observatory site. We thank Mark Wieringa and Hayley Bignall for their help with various aspects of the ATCA data, Ian Browne for his usual words of wisdom and the referee, Neal Jackson, for his report.

\section*{Data Availability}

The data underlying this article will be shared on reasonable request to the corresponding author.



\bibliographystyle{mnras}
\bibliography{lensing}

\begin{thebibliography}{}
\makeatletter
\relax
\def\mn@urlcharsother{\let\do\@makeother \do\$\do\&\do\#\do\^\do\_\do\%\do\~}
\def\mn@doi{\begingroup\mn@urlcharsother \@ifnextchar [ {\mn@doi@}
  {\mn@doi@[]}}
\def\mn@doi@[#1]#2{\def\@tempa{#1}\ifx\@tempa\@empty \href
  {http://dx.doi.org/#2} {doi:#2}\else \href {http://dx.doi.org/#2} {#1}\fi
  \endgroup}
\def\mn@eprint#1#2{\mn@eprint@#1:#2::\@nil}
\def\mn@eprint@arXiv#1{\href {http://arxiv.org/abs/#1} {{\tt arXiv:#1}}}
\def\mn@eprint@dblp#1{\href {http://dblp.uni-trier.de/rec/bibtex/#1.xml}
  {dblp:#1}}
\def\mn@eprint@#1:#2:#3:#4\@nil{\def\@tempa {#1}\def\@tempb {#2}\def\@tempc
  {#3}\ifx \@tempc \@empty \let \@tempc \@tempb \let \@tempb \@tempa \fi \ifx
  \@tempb \@empty \def\@tempb {arXiv}\fi \@ifundefined
  {mn@eprint@\@tempb}{\@tempb:\@tempc}{\expandafter \expandafter \csname
  mn@eprint@\@tempb\endcsname \expandafter{\@tempc}}}

\bibitem[\protect\citeauthoryear{{Abdo} et~al.,}{{Abdo} et~al.}{2015}]{abdo15}
{Abdo} A.~A.,  et~al., 2015, \mn@doi [\apj] {10.1088/0004-637X/799/2/143},
  \href {https://ui.adsabs.harvard.edu/abs/2015ApJ...799..143A} {799, 143}

\bibitem[\protect\citeauthoryear{{Abhir}, {Prince}, {Joseph}, {Bose}  \&
  {Gupta}}{{Abhir} et~al.}{2021}]{abhir21}
{Abhir} J.,  {Prince} R.,  {Joseph} J.,  {Bose} D.,   {Gupta} N.,  2021,
  \mn@doi [\apj] {10.3847/1538-4357/abfd33}, \href
  {https://ui.adsabs.harvard.edu/abs/2021ApJ...915...26A} {915, 26}

\bibitem[\protect\citeauthoryear{{Barnacka}, {Glicenstein}  \&
  {Moudden}}{{Barnacka} et~al.}{2011}]{barnacka11}
{Barnacka} A.,  {Glicenstein} J.~F.,   {Moudden} Y.,  2011, \mn@doi [\aap]
  {10.1051/0004-6361/201016175}, \href
  {https://ui.adsabs.harvard.edu/abs/2011A&A...528L...3B} {528, L3}

\bibitem[\protect\citeauthoryear{{Barnacka}, {Geller}, {Dell'Antonio}  \&
  {Benbow}}{{Barnacka} et~al.}{2015}]{barnacka15}
{Barnacka} A.,  {Geller} M.~J.,  {Dell'Antonio} I.~P.,   {Benbow} W.,  2015,
  \mn@doi [\apj] {10.1088/0004-637X/809/1/100}, \href
  {https://ui.adsabs.harvard.edu/abs/2015ApJ...809..100B} {809, 100}

\bibitem[\protect\citeauthoryear{{Biggs}}{{Biggs}}{2021}]{biggs21}
{Biggs} A.~D.,  2021, \mn@doi [\mnras] {10.1093/mnras/stab1444}, \href
  {https://ui.adsabs.harvard.edu/abs/2021MNRAS.505.2610B} {505, 2610}

\bibitem[\protect\citeauthoryear{{Biggs} \& {Browne}}{{Biggs} \&
  {Browne}}{2018}]{biggs18}
{Biggs} A.~D.,  {Browne} I.~W.~A.,  2018, \mn@doi [\mnras]
  {10.1093/mnras/sty565}, \href
  {https://ui.adsabs.harvard.edu/#abs/2018MNRAS.476.5393B} {476, 5393}

\bibitem[\protect\citeauthoryear{{CASA Team} et~al.,}{{CASA Team}
  et~al.}{2022}]{casa22}
{CASA Team} et~al., 2022, \mn@doi [\pasp] {10.1088/1538-3873/ac9642}, \href
  {https://ui.adsabs.harvard.edu/abs/2022PASP..134k4501C} {134, 114501}

\bibitem[\protect\citeauthoryear{{Chengalur}, {de Bruyn}  \&
  {Narasimha}}{{Chengalur} et~al.}{1999}]{chengalur99}
{Chengalur} J.~N.,  {de Bruyn} A.~G.,   {Narasimha} D.,  1999, \aap, \href
  {http://adsabs.harvard.edu/abs/1999A%26A...343L..79C} {343, L79}

\bibitem[\protect\citeauthoryear{{Cheung} et~al.,}{{Cheung}
  et~al.}{2014}]{cheung14}
{Cheung} C.~C.,  et~al., 2014, \mn@doi [\apjl] {10.1088/2041-8205/782/2/L14},
  \href {http://esoads.eso.org/abs/2014ApJ...782L..14C} {782, L14}

\bibitem[\protect\citeauthoryear{{Combes} et~al.,}{{Combes}
  et~al.}{2021}]{combes21}
{Combes} F.,  et~al., 2021, \mn@doi [\aap] {10.1051/0004-6361/202040167}, \href
  {https://ui.adsabs.harvard.edu/abs/2021A&A...648A.116C} {648, A116}

\bibitem[\protect\citeauthoryear{{Courbin}, {Meylan}, {Kneib}  \&
  {Lidman}}{{Courbin} et~al.}{2002}]{courbin02}
{Courbin} F.,  {Meylan} G.,  {Kneib} J.~P.,   {Lidman} C.,  2002, \mn@doi
  [\apj] {10.1086/341261}, \href
  {https://ui.adsabs.harvard.edu/abs/2002ApJ...575...95C} {575, 95}

\bibitem[\protect\citeauthoryear{{Dallacasa}, {Bondi}, {Alef}  \&
  {Mantovani}}{{Dallacasa} et~al.}{1998}]{dallacasa98}
{Dallacasa} D.,  {Bondi} M.,  {Alef} W.,   {Mantovani} F.,  1998, \mn@doi
  [\aaps] {10.1051/aas:1998183}, \href
  {https://ui.adsabs.harvard.edu/abs/1998A&AS..129..219D} {129, 219}

\bibitem[\protect\citeauthoryear{{Dallacasa}, {Orienti}, {Fanti}, {Fanti}  \&
  {Stanghellini}}{{Dallacasa} et~al.}{2013}]{dallacasa13}
{Dallacasa} D.,  {Orienti} M.,  {Fanti} C.,  {Fanti} R.,   {Stanghellini} C.,
  2013, \mn@doi [\mnras] {10.1093/mnras/stt710}, \href
  {https://ui.adsabs.harvard.edu/abs/2013MNRAS.433..147D} {433, 147}

\bibitem[\protect\citeauthoryear{{Ellithorpe}, {Kochanek}  \&
  {Hewitt}}{{Ellithorpe} et~al.}{1996}]{ellithorpe96}
{Ellithorpe} J.~D.,  {Kochanek} C.~S.,   {Hewitt} J.~N.,  1996, \mn@doi [\apj]
  {10.1086/177346}, \href
  {https://ui.adsabs.harvard.edu/abs/1996ApJ...464..556E} {464, 556}

\bibitem[\protect\citeauthoryear{{Fassnacht} \& {Taylor}}{{Fassnacht} \&
  {Taylor}}{2001}]{fassnacht01}
{Fassnacht} C.~D.,  {Taylor} G.~B.,  2001, \mn@doi [\aj] {10.1086/322112},
  \href {http://adsabs.harvard.edu/abs/2001AJ....122.1661F} {122, 1661}

\bibitem[\protect\citeauthoryear{{Fassnacht}, {Xanthopoulos}, {Koopmans}  \&
  {Rusin}}{{Fassnacht} et~al.}{2002}]{fassnacht02}
{Fassnacht} C.~D.,  {Xanthopoulos} E.,  {Koopmans} L.~V.~E.,   {Rusin} D.,
  2002, \mn@doi [\apj] {10.1086/344368}, \href
  {http://adsabs.harvard.edu/abs/2002ApJ...581..823F} {581, 823}

\bibitem[\protect\citeauthoryear{{Garrett}, {Nair}, {Porcas}  \&
  {Patnaik}}{{Garrett} et~al.}{1997}]{garrett97}
{Garrett} M.~A.,  {Nair} S.,  {Porcas} R.~W.,   {Patnaik} A.~R.,  1997, \mn@doi
  [Vistas in Astronomy] {10.1016/S0083-6656(97)00020-2}, \href
  {https://ui.adsabs.harvard.edu/abs/1997VA.....41..281G} {41, 281}

\bibitem[\protect\citeauthoryear{{Greisen}}{{Greisen}}{2003}]{greisen03}
{Greisen} E.~W.,  2003, in {Heck} A.,  ed.,  Astrophysics and Space Science
  Library Vol. 285, Information Handling in Astronomy - Historical Vistas.
  p.~109, \mn@doi{10.1007/0-306-48080-8_7}

\bibitem[\protect\citeauthoryear{{Guirado}, {Jones}, {Lara}, {Marcaide},
  {Preston}, {Rao}  \& {Sherwood}}{{Guirado} et~al.}{1999}]{guirado99}
{Guirado} J.~C.,  {Jones} D.~L.,  {Lara} L.,  {Marcaide} J.~M.,  {Preston}
  R.~A.,  {Rao} A.~P.,   {Sherwood} W.~A.,  1999, \aap, \href
  {http://adsabs.harvard.edu/abs/1999A%26A...346..392G} {346, 392}

\bibitem[\protect\citeauthoryear{{Jauncey} et~al.,}{{Jauncey}
  et~al.}{1991}]{jauncey91}
{Jauncey} D.~L.,  et~al., 1991, \mn@doi [\nat] {10.1038/352132a0}, \href
  {https://ui.adsabs.harvard.edu/abs/1991Natur.352..132J} {352, 132}

\bibitem[\protect\citeauthoryear{{Jones} et~al.,}{{Jones}
  et~al.}{1996}]{jones96}
{Jones} D.~L.,  et~al., 1996, \mn@doi [\apjl] {10.1086/310292}, \href
  {http://adsabs.harvard.edu/abs/1996ApJ...470L..23J} {470, L23}

\bibitem[\protect\citeauthoryear{{Kochanek} \& {Narayan}}{{Kochanek} \&
  {Narayan}}{1992}]{kochanek92}
{Kochanek} C.~S.,  {Narayan} R.,  1992, \mn@doi [\apj] {10.1086/172078}, \href
  {http://esoads.eso.org/abs/1992ApJ...401..461K} {401, 461}

\bibitem[\protect\citeauthoryear{{Koopmans} \& {de Bruyn}}{{Koopmans} \& {de
  Bruyn}}{2000}]{koopmans00b}
{Koopmans} L.~V.~E.,  {de Bruyn} A.~G.,  2000, \aap, \href
  {http://adsabs.harvard.edu/abs/2000A%26A...358..793K} {358, 793}

\bibitem[\protect\citeauthoryear{{Leh{\'a}r} et~al.,}{{Leh{\'a}r}
  et~al.}{2000}]{lehar00}
{Leh{\'a}r} J.,  et~al., 2000, \mn@doi [\apj] {10.1086/308963}, \href
  {http://adsabs.harvard.edu/abs/2000ApJ...536..584L} {536, 584}

\bibitem[\protect\citeauthoryear{{Lidman}, {Courbin}, {Meylan}, {Broadhurst},
  {Frye}  \& {Welch}}{{Lidman} et~al.}{1999}]{lidman99}
{Lidman} C.,  {Courbin} F.,  {Meylan} G.,  {Broadhurst} T.,  {Frye} B.,
  {Welch} W.~J.~W.,  1999, \mn@doi [\apjl] {10.1086/311949}, \href
  {https://ui.adsabs.harvard.edu/abs/1999ApJ...514L..57L} {514, L57}

\bibitem[\protect\citeauthoryear{{Lovell} et~al.,}{{Lovell}
  et~al.}{1996}]{lovell96}
{Lovell} J.~E.~J.,  et~al., 1996, \mn@doi [\apjl] {10.1086/310353}, \href
  {https://ui.adsabs.harvard.edu/abs/1996ApJ...472L...5L} {472, L5}

\bibitem[\protect\citeauthoryear{{Lovell}, {Jauncey}, {Reynolds}, {Wieringa},
  {King}, {Tzioumis}, {McCulloch}  \& {Edwards}}{{Lovell}
  et~al.}{1998}]{lovell98}
{Lovell} J.~E.~J.,  {Jauncey} D.~L.,  {Reynolds} J.~E.,  {Wieringa} M.~H.,
  {King} E.~A.,  {Tzioumis} A.~K.,  {McCulloch} P.~M.,   {Edwards} P.~G.,
  1998, \mn@doi [\apj] {10.1086/311723}, \href
  {https://ui.adsabs.harvard.edu/\#abs/1998ApJ...508L..51L} {508, L51}

\bibitem[\protect\citeauthoryear{{Meylan}, {Courbin}, {Lidman}, {Kneib}  \&
  {Tacconi-Garman}}{{Meylan} et~al.}{2005}]{meylan05}
{Meylan} G.,  {Courbin} F.,  {Lidman} C.,  {Kneib} J.~P.,   {Tacconi-Garman}
  L.~E.,  2005, \mn@doi [\aap] {10.1051/0004-6361:200500145}, \href
  {https://ui.adsabs.harvard.edu/abs/2005A&A...438L..37M} {438, L37}

\bibitem[\protect\citeauthoryear{{Muller} \& {Gu{\'e}lin}}{{Muller} \&
  {Gu{\'e}lin}}{2008}]{muller08}
{Muller} S.,  {Gu{\'e}lin} M.,  2008, \mn@doi [\aap]
  {10.1051/0004-6361:200810392}, \href
  {https://ui.adsabs.harvard.edu/abs/2008A&A...491..739M} {491, 739}

\bibitem[\protect\citeauthoryear{{Muller}, {Jaswanth}, {Horellou}  \&
  {Mart{\'\i}-Vidal}}{{Muller} et~al.}{2020}]{muller20}
{Muller} S.,  {Jaswanth} S.,  {Horellou} C.,   {Mart{\'\i}-Vidal} I.,  2020,
  \mn@doi [\aap] {10.1051/0004-6361/202038978}, \href
  {https://ui.adsabs.harvard.edu/abs/2020A&A...641L...2M} {641, L2}

\bibitem[\protect\citeauthoryear{{Muller} et~al.,}{{Muller}
  et~al.}{2023}]{muller23}
{Muller} S.,  et~al., 2023, \mn@doi [\aap] {10.1051/0004-6361/202245768}, \href
  {https://ui.adsabs.harvard.edu/abs/2023A&A...674A.101M} {674, A101}

\bibitem[\protect\citeauthoryear{{Nair}, {Narasimha}  \& {Rao}}{{Nair}
  et~al.}{1993}]{nair93}
{Nair} S.,  {Narasimha} D.,   {Rao} A.~P.,  1993, \mn@doi [\apj]
  {10.1086/172491}, \href
  {https://ui.adsabs.harvard.edu/abs/1993ApJ...407...46N} {407, 46}

\bibitem[\protect\citeauthoryear{{Narayan}}{{Narayan}}{1992}]{narayan92}
{Narayan} R.,  1992, \mn@doi [Phil. Trans. Roy. Soc.] {10.1098/rsta.1992.0090},
  \href {https://ui.adsabs.harvard.edu/abs/1992RSPTA.341..151N} {341, 151}

\bibitem[\protect\citeauthoryear{{Nightingale} \& {Dye}}{{Nightingale} \&
  {Dye}}{2015}]{nightingale15}
{Nightingale} J.~W.,  {Dye} S.,  2015, \mn@doi [\mnras]
  {10.1093/mnras/stv1455}, \href
  {https://ui.adsabs.harvard.edu/abs/2015MNRAS.452.2940N} {452, 2940}

\bibitem[\protect\citeauthoryear{{Nightingale}, {Dye}  \&
  {Massey}}{{Nightingale} et~al.}{2018}]{nightingale18}
{Nightingale} J.~W.,  {Dye} S.,   {Massey} R.~J.,  2018, \mn@doi [\mnras]
  {10.1093/mnras/sty1264}, \href
  {https://ui.adsabs.harvard.edu/abs/2018MNRAS.478.4738N} {478, 4738}

\bibitem[\protect\citeauthoryear{{Nightingale} et~al.,}{{Nightingale}
  et~al.}{2021}]{nightingale21}
{Nightingale} J.,  et~al., 2021, \mn@doi [The Journal of Open Source Software]
  {10.21105/joss.02825}, \href
  {https://ui.adsabs.harvard.edu/abs/2021JOSS....6.2825N} {6, 2825}

\bibitem[\protect\citeauthoryear{{Peck} \& {Taylor}}{{Peck} \&
  {Taylor}}{2000}]{peck00}
{Peck} A.,  {Taylor} G.~B.,  2000, in {Conway} J.~E.,  {Polatidis} A.~G.,
  {Booth} R.~S.,   {Pihlstr{\"o}m} Y.~M.,  eds, Proc. 5th EVN Symp. Onsala
  Space Observatory, Onsala, p.~95 (\mn@eprint {} {astro-ph/0009372})

\bibitem[\protect\citeauthoryear{{Peirson} et~al.,}{{Peirson}
  et~al.}{2022}]{peirson22}
{Peirson} A.~L.,  et~al., 2022, \mn@doi [\apj] {10.3847/1538-4357/ac469e},
  \href {https://ui.adsabs.harvard.edu/abs/2022ApJ...927...24P} {927, 24}

\bibitem[\protect\citeauthoryear{{Pelt}, {Kayser}, {Refsdal}  \&
  {Schramm}}{{Pelt} et~al.}{1996}]{pelt96}
{Pelt} J.,  {Kayser} R.,  {Refsdal} S.,   {Schramm} T.,  1996, \aap, \href
  {http://esoads.eso.org/abs/1996A%26A...305...97P} {305, 97}

\bibitem[\protect\citeauthoryear{{Pramesh Rao} \& {Subrahmanyan}}{{Pramesh Rao}
  \& {Subrahmanyan}}{1988}]{prameshrao88}
{Pramesh Rao} A.,  {Subrahmanyan} R.,  1988, \mn@doi [\mnras]
  {10.1093/mnras/231.2.229}, \href
  {https://ui.adsabs.harvard.edu/abs/1988MNRAS.231..229P} {231, 229}

\bibitem[\protect\citeauthoryear{{Rickett}}{{Rickett}}{1990}]{rickett90}
{Rickett} B.~J.,  1990, \mn@doi [\araa] {10.1146/annurev.aa.28.090190.003021},
  \href {https://ui.adsabs.harvard.edu/abs/1990ARA&A..28..561R} {28, 561}

\bibitem[\protect\citeauthoryear{{Shepherd}}{{Shepherd}}{1997}]{shepherd97}
{Shepherd} M.~C.,  1997, in {Hunt} G.,  {Payne} H.,  eds, ASP. Conf. Ser. Vol.
  125, Astronomical Data Analysis Software and Systems VI. Astron. Soc. Pac.,
  San Francisco, p.~77

\bibitem[\protect\citeauthoryear{{Stanghellini}, {Bondi}, {Dallacasa}, {O'Dea},
  {Baum}, {Fanti}  \& {Fanti}}{{Stanghellini} et~al.}{1997}]{stanghellini97}
{Stanghellini} C.,  {Bondi} M.,  {Dallacasa} D.,  {O'Dea} C.~P.,  {Baum} S.~A.,
   {Fanti} R.,   {Fanti} C.,  1997, \aap, \href
  {https://ui.adsabs.harvard.edu/abs/1997A&A...318..376S} {318, 376}

\bibitem[\protect\citeauthoryear{{Subrahmanyan}, {Narasimha}, {Pramesh Rao}  \&
  {Swarup}}{{Subrahmanyan} et~al.}{1990}]{subrahmanyan90}
{Subrahmanyan} R.,  {Narasimha} D.,  {Pramesh Rao} A.,   {Swarup} G.,  1990,
  \mnras, \href {https://ui.adsabs.harvard.edu/abs/1990MNRAS.246..263S} {246,
  263}

\bibitem[\protect\citeauthoryear{{Tingay}, {Jauncey}, {King}, {Tzioumis},
  {Lovell}  \& {Edwards}}{{Tingay} et~al.}{2003}]{tingay03}
{Tingay} S.~J.,  {Jauncey} D.~L.,  {King} E.~A.,  {Tzioumis} A.~K.,  {Lovell}
  J. E.~J.,   {Edwards} P.~G.,  2003, \mn@doi [\pasj] {10.1093/pasj/55.2.351},
  \href {https://ui.adsabs.harvard.edu/abs/2003PASJ...55..351T} {55, 351}

\bibitem[\protect\citeauthoryear{{Vedantham} et~al.,}{{Vedantham}
  et~al.}{2017}]{vedantham17}
{Vedantham} H.~K.,  et~al., 2017, \mn@doi [\apj] {10.3847/1538-4357/aa745c},
  \href {https://ui.adsabs.harvard.edu/abs/2017ApJ...845...89V} {845, 89}

\bibitem[\protect\citeauthoryear{{Verheijen}, {Carilli}  \& {Yun}}{{Verheijen}
  et~al.}{2001}]{verheijen01}
{Verheijen} M.~A.~W.,  {Carilli} C.~L.,   {Yun} M.~S.,  2001, in {Hibbard}
  J.~E.,  {Rupen} M.,   {van Gorkom} J.~H.,  eds,  Astronomical Society of the
  Pacific Conference Series Vol. 240, Gas and Galaxy Evolution. p.~69

\bibitem[\protect\citeauthoryear{{Walker}}{{Walker}}{1998}]{walker98}
{Walker} M.~A.,  1998, \mn@doi [\mnras] {10.1046/j.1365-8711.1998.01238.x},
  \href {https://ui.adsabs.harvard.edu/abs/1998MNRAS.294..307W} {294, 307}

\bibitem[\protect\citeauthoryear{{Walker}}{{Walker}}{2001}]{walker01}
{Walker} M.~A.,  2001, \mn@doi [\mnras] {10.1046/j.1365-8711.2001.04104.x},
  \href {https://ui.adsabs.harvard.edu/abs/2001MNRAS.321..176W} {321, 176}

\bibitem[\protect\citeauthoryear{{Wardle} \& {Kronberg}}{{Wardle} \&
  {Kronberg}}{1974}]{wardle74}
{Wardle} J.~F.~C.,  {Kronberg} P.~P.,  1974, \mn@doi [\apj] {10.1086/153240},
  \href {https://ui.adsabs.harvard.edu/#abs/1974ApJ...194..249W} {194, 249}

\bibitem[\protect\citeauthoryear{{Warren} \& {Dye}}{{Warren} \&
  {Dye}}{2003}]{warren03}
{Warren} S.~J.,  {Dye} S.,  2003, \mn@doi [\apj] {10.1086/375132}, \href
  {http://adsabs.harvard.edu/abs/2003ApJ...590..673W} {590, 673}

\bibitem[\protect\citeauthoryear{{Wiklind} \& {Combes}}{{Wiklind} \&
  {Combes}}{1996}]{wiklind96}
{Wiklind} T.,  {Combes} F.,  1996, \mn@doi [\nat] {10.1038/379139a0}, \href
  {https://ui.adsabs.harvard.edu/abs/1996Natur.379..139W} {379, 139}

\bibitem[\protect\citeauthoryear{{Wiklind} \& {Combes}}{{Wiklind} \&
  {Combes}}{2001}]{wiklind01}
{Wiklind} T.,  {Combes} F.,  2001, in {Brainerd} T.~G.,  {Kochanek} C.~S.,
  eds,  Astronomical Society of the Pacific Conference Series Vol. 237,
  Gravitational Lensing: Recent Progress and Future Goals. p.~155 (\mn@eprint
  {arXiv} {astro-ph/9909314}), \mn@doi{10.48550/arXiv.astro-ph/9909314}

\bibitem[\protect\citeauthoryear{{Winn}, {Hewitt}  \& {Schechter}}{{Winn}
  et~al.}{2001}]{winn01b}
{Winn} J.~N.,  {Hewitt} J.~N.,   {Schechter} P.~L.,  2001, in {Brainerd} T.~G.,
   {Kochanek} C.~S.,  eds,  Astronomical Society of the Pacific Conference
  Series Vol. 237, Gravitational Lensing: Recent Progress and Future Goals.
  p.~61 (\mn@eprint {arXiv} {astro-ph/9909335}),
  \mn@doi{10.48550/arXiv.astro-ph/9909335}

\bibitem[\protect\citeauthoryear{{Winn}, {Kochanek}, {McLeod}, {Falco}, {Impey}
   \& {Rix}}{{Winn} et~al.}{2002}]{winn02a}
{Winn} J.~N.,  {Kochanek} C.~S.,  {McLeod} B.~A.,  {Falco} E.~E.,  {Impey}
  C.~D.,   {Rix} H.-W.,  2002, \mn@doi [\apj] {10.1086/341265}, \href
  {https://ui.adsabs.harvard.edu/abs/2002ApJ...575..103W} {575, 103}

\bibitem[\protect\citeauthoryear{{Winn} et~al.,}{{Winn} et~al.}{2004}]{winn04a}
{Winn} J.~N.,  et~al., 2004, \mn@doi [\aj] {10.1086/425881}, \href
  {https://ui.adsabs.harvard.edu/abs/2004AJ....128.2696W} {128, 2696}

\bibitem[\protect\citeauthoryear{{Wucknitz}}{{Wucknitz}}{2004}]{wucknitz04a}
{Wucknitz} O.,  2004, \mn@doi [\mnras] {10.1111/j.1365-2966.2004.07513.x},
  \href {http://esoads.eso.org/abs/2004MNRAS.349....1W} {349, 1}

\bibitem[\protect\citeauthoryear{{van Ommen}, {Jones}, {Preston}  \&
  {Jauncey}}{{van Ommen} et~al.}{1995}]{vanommen95}
{van Ommen} T.~D.,  {Jones} D.~L.,  {Preston} R.~A.,   {Jauncey} D.~L.,  1995,
  \mn@doi [\apj] {10.1086/175630}, \href
  {https://ui.adsabs.harvard.edu/abs/1995ApJ...444..561V} {444, 561}

\makeatother
\end{thebibliography}


\bsp	
\label{lastpage}
\end{document}